\documentclass[twocolumn]{el-author}
\usepackage{epsfig,graphicx,epstopdf}
\date{}

\begin{document}

\bibliographystyle{ieee}

\title{A Generalized Correlation Index for Quantifying
Signal Morphological Similarity}
\author{A. Olenko, K. T. Wong, H. Mir, and H. Al-Nashash}

\abstract{
In biomedical applications, the similarity between a signal measured from an injured subject and a reference signal measured from a normal subject can be used to quantify the injury severity. This paper proposes a generalization of the adaptive signed correlation index (ASCI)  to account for specific signal features of interest and extend the trichotomization of conventional ASCI to an arbitrary number of levels. In the context of spinal cord injury assessment, a computational example is presented to illustrate the enhanced resolution of the proposed measure and its ability to offer a more refined measure of the level of injury.}

\maketitle

\section{Introduction}
\label{sec1}

Of special salience to some biomedical applications is the comparison between  a signal measured from an injured subject and a  reference signal measured from a normal subject. For example, an objective quantitative assessment method is useful to enable the monitoring of spinal cord injury (SCI) recovery and rehabilitation and to assess the effectiveness of any possible therapeutic mechanisms.  A powerful technique used in SCI studies is  the Somatosensory Evoked Potential (SEP), which is the cortical signal recorded in response to sensory stimulation. For thoracic-level SCI, the  SEP recorded in response to forelimb stimulation corresponds to the reference signal, while the SEP recorded in response to hindlimb stimulation corresponds to the injured signal.
These signals differ by an amplitude-offset which may vary over the entire time duration of interest, and the difference in the amplitude values of the signals is useful in developing
a quantitative measure of the severity of the injury \cite{AlNashashBET0809},\cite{MirICBBE2010}.  Therefore, it is of clinical interest to quantify the proximity of the corresponding amplitude values of the signals. 
\begin{figure}[!ht]
 \centering
 \scalebox{0.4}{\includegraphics{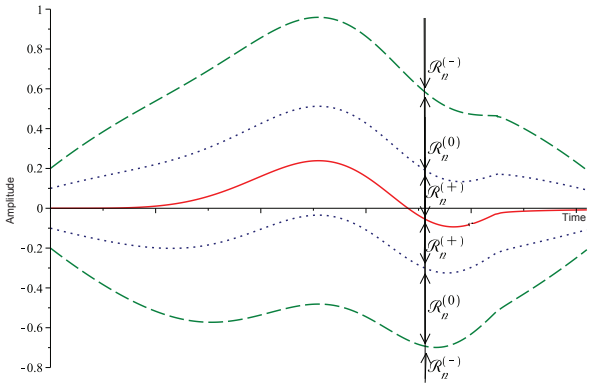}}
\caption{Test signal (solid) and vertical shifts of  reference signal (dashed) used to define  boundaries of  subranges ${\cal R}_n^{(+)}$, ${\cal R}_n^{(0)}$, and ${\cal R}_n^{(-)}$. }
\label{fig00}
\end{figure}

As an example, consider the trichotomization problem  \cite{LianSP0210} illustrated in Figure \ref{fig00}.
A test signal defined as an $N$-tuple of real-valued numbers, ${\bf x} := (x_1, \cdots, x_N)$, is to have its $n$th value ($x_n$) classified, for each $n$,  into one of three disjoint subranges (each possibly non-contiguous): 
\begin{eqnarray} 
{\cal R}_n^{(+)} 
&:=& 
\left(c_n - R_n^{(+)}, ~ c_n + R_n^{(+)} \right), \nonumber \\
{\cal R}_n^{(0)} 
&:=& 
\left(c_n - R_n^{(+)} - R_n^{(0)}, ~ c_n - R_n^{(+)} \right)  \nonumber \\
&& 
~ \cup ~   \left(c_n + R_n^{(+)}, ~ c_n + R_n^{(+)} + R_n^{(0)}\right),  \nonumber \\
{\cal R}_n^{(-)} 
&:=& 
\left(-\infty, ~ c_n - R_n^{(+)} - R_n^{(0)} \right)  \nonumber \\
&& 
~ \cup ~   \left( c_n + R_n^{(+)} + R_n^{(0)}, ~ \infty \right), 
\end{eqnarray}
where ${\bf c} := (c_1, \cdots, c_N)$ represents a reference signal in discrete time.
The subranges of ${\cal R}_n^{(+)}$, ${\cal R}_n^{(0)}$, ${\cal R}_n^{(-)}$ may vary with the time-index $n$. The $n$th element of ${\bf x}$ is trichotomized as 
\begin{eqnarray} \label{eq1}
T(x_n)
&:=&
\left\{ \begin{array}{rl}
 1, &\mbox{if } x_n \in {\cal R}_n^{(+)}; \\
 0, &\mbox{if } x_n \in {\cal R}_n^{(0)}; \\
-1, &\mbox{if } x_n \in {\cal R}_n^{(-)}. 
\end{array} \right.
\end{eqnarray}
It is in terms of this trichotomization that 
two measured signals (${\bf x}$ and ${\bf y} := (y_1, \cdots, y_N)$)  
are to have their morphological similarity quantified. Towards this end, the 
adaptive signed correlation index (ASCI) has been proposed in  \cite{LianSP0210}, where
\begin{eqnarray}\label{eq2}
{\rm ASCI}({\bf x},{\bf y})
&:=&
\frac{1}{N}\sum_{n=1}^N \left[T(x_n)\otimes T(y_n)\right],
\end{eqnarray}
and
\begin{eqnarray}\label{eq3}
T(x_n)\otimes T(y_n)
&:=&
\left\{ \begin{array}{rl}
 1, &\mbox{if } T(x_n)=T(y_n), \\
-1, &\mbox{if } T(x_n)\cdot T(y_n)=-1, \\
 0, &\mbox{otherwise,}
\end{array} \right.  
\label{ASCI-metric}
\end{eqnarray}
for each discrete-time index $n$. ASCI 
uses a balanced ternary   extension of the boolean algebraic operator XNOR  that indicates binary logic equality. As such, the closeness between two signals
${\bf x}$ and ${\bf y}$ is defined 
(i) in terms of  the relative magnitudes of their amplitude shifts $T(x_n)$ and $T(y_n)$, at any {\em specific} $n$, and 
(ii) with respect to the similarity between their shapes, i.e. $T(x_n) \otimes T(y_n)$ {\em over all} $n$. Such a measure differs from the conventional similarity  measured by Pearson's 
correlation coefficient, which is invariant of any affine transformation. Indeed, the ASCI has been shown in \cite{LianSP0210} for this reason to be  advantageous (over Pearson's correlation coefficient) in quantifying the  similarity between clinically measured neurological signals.

This paper proposes a generalization of the ASCI to overcome its resolution limitations and offer a more refined estimate of signal morphological similarity. The proposed generalization differs from a straight-forward increase in the number of quantization levels since 
not only one scalar/vector is measured at each time instant (as in customary quantization), but a nonlinear and non-contiguously defined relationship between two vectors of measurements with time-varying segmentation boundaries is accounted for. The process of generalizing the three-value logic operator in (\ref{eq3}) is not obvious, since an appropriate logic operator must be developed to operate on a pair of such generalized  multiple-level variables.  
Moreover, since $K$  is finite, this subset does not form a linear subspace in  the sequence space, and care is needed to ensure that  the proposed generalization results in a distance metric that also preserves the important properties of the original ASCI.

\section{Generalization of ASCI}
\label{sec4}

To prepare for the subsequent generalization, 
the ASCI will first be reformulated and re-interpreted in terms of the distances between the trichotomized values, instead of the signed product of (\ref{ASCI-metric}). From (\ref{eq1}),
\begin{eqnarray}
T(x_n)-T(y_n)
&=&
\left\{ \begin{array}{rl}
    0, &\mbox{if }  T(x_n) = T(y_n) \\
\pm 1, &\mbox{if }  T(x_n) \cdot T(y_n)=0 \\
\pm 2, &\mbox{if }  T(x_n) =-T(y_n) \not=0.
\end{array} \right.
\end{eqnarray}
Therefore, re-write
\begin{eqnarray}
T(x_n) \otimes T(y_n)
&=&1- \left|T(x_n) - T(y_n) \right|,\\
{\rm ASCI}({\bf x},{\bf y}) 
&=& 
1-\frac{1}{N}\sum_{n=1}^N \left|T(x_n) - T(y_n) \right|.
\end{eqnarray}
Next, assign a new set of numerical values to the trichotomization:
\begin{eqnarray}
\label{tilt}
\tilde{T}(x_n)
&:=&
\left\{ \begin{array}{rl} 
1, &\mbox{if }x_n \in {\cal R}_n^{(+)} \\
2, &\mbox{if }x_n \in {\cal R}_n^{(0)} \\
3, &\mbox{if }x_n \in {\cal R}_n^{(-)} . 
\end{array} \right.
\end{eqnarray}
Since $\tilde{T}(\cdot)$ does not require any symmetry in its value with respect to zero, it will be more convenient to use than $T(\cdot)$. Using this notation, (\ref{eq3}) and (\ref{eq2}) respectively become  
\begin{eqnarray}
T(x_n) \otimes T(y_n)
&=&
1- \left|\tilde{T}(x_n) - \tilde{T}(y_n) \right|, \\
\label{asci}
{\rm ASCI}({\bf x},{\bf y})
&=&
1-\frac{1}{N}\sum_{n=1}^N  \left| \tilde{T}(x_n) - \tilde{T}(y_n) \right|.
\end{eqnarray}
This reformulation will lend itself naturally to the subsequent poly-partite generalization of the ASCI.

A finer partition of the dynamic range of the signals 
will enable the capture of small amplitude differences between them. 
Hence, this paper generalizes the original  {\em three} disjoint subranges to $K$ subranges $\left\{ {\cal R}_n^{(k)}, ~~ k=1,\cdots,K \right\}$, where ${\cal R}_n =\cup_{k=1}^K {\cal R}_n^{(k)}$, and ${\cal R}_n^{(k)} \cap {\cal R}_n^{(j)} = \phi, \forall k \neq j$.
A sharpened resolution in a generalized ASCI  requires more gradations of value in the variable $T(\cdot)$. A suitable logic operator must therefore be defined to handle such generalized multiple-level variables. 

As such, generalize the original indicator function $T(x_n)$ to
\begin{eqnarray}\label{t1}
\tilde{T}(x_n)
&:=&
k,\quad \mbox{if } x_n\in {\cal R}_n^{(k)}, ~~ k=1,2,\cdots,K.
\end{eqnarray} 
Thus, $\tilde{T}(x_n) - \tilde{T}(y_n) \in \{1-K,\cdots,K-1\}.$
Further define
\begin{eqnarray}
\tilde{T}(x_n) \otimes \tilde{T}(y_n)
&:=&
1-\frac{2}{K-1} \left| \tilde{T}(x_n) - \tilde{T}(y_n) \right|,  \\
{\rm ASCI}_K({\bf x},{\bf y})
&:=&
1-\frac{2}{N(K-1)}\sum_{n=1}^N  \left|\tilde{T}(x_n) - \tilde{T}(y_n) \right|.
   \label{asci1}
\end{eqnarray}
Note that at $K=3$, ASCI$_3({\bf x},{\bf y})$ would coincide with the conventional ASCI$({\bf x},{\bf y})$. Furthermore, it can be shown that the newly defined ASCI$_K({\bf x},{\bf y})$ satisfies the usual properties of normalized indices:
\begin{itemize}
\item  ASCI$_K({\bf x},{\bf y}) \in [-1,1]$.
\item  If $\tilde{T}(x_n)=\tilde{T}(y_n)$ for all $n=1,\cdots,N$, then ${\rm ASCI}_K({\bf x},{\bf y}) =1$.
\item  If $\tilde{T}(x_n)=K$ and $\tilde{T}(y_n)=1, \forall n=1,\cdots,N$ (the case of maximum possible amplitude differences), then ${\rm ASCI}_K({\bf x},{\bf y}) = -1$.
\item  ASCI$_K({\bf x},{\bf y}) =  {\rm ASCI}_K({\bf y},{\bf x})$.
\item It is translationally invariant, in the sense that 
ASCI$_K({\bf x},{\bf y}) = {\rm ASCI}_K({\bf z},{\bf w})$, if 
$\left|\tilde{T}(x_n) -\tilde{T}(y_n) \right| ~=~ 
 \left|\tilde{T}(z_n) -\tilde{T}(w_n) \right|$, $\forall n$.
\end{itemize}

\section{Computational Example} \label{sec5}

\begin{figure}[!ht]
 \centering
 \scalebox{.5}{\includegraphics{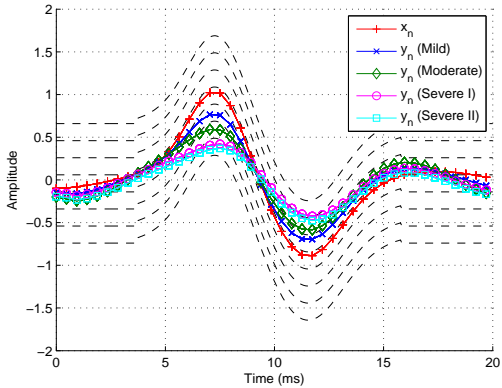}}
 \caption{Injury SEP signals (solid) and  shifts of reference SEP signal (dashed) for $K=7$.}
\label{k7}
\end{figure}

\begin{figure}[!ht]
 \centering
 \scalebox{.55}{\includegraphics{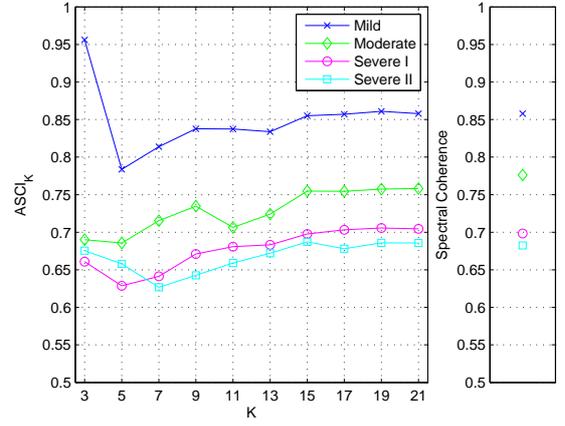}}
 \caption{Dependence of generalized ASCI values on $K$.}
\label{ascik}
\end{figure}

Consider again the issue of SCI assessment.  A set of  SEP signals based on those used in \cite{MirICBBE2010, vipin2} is shown in Figure \ref{k7}, where $x_n$ is the forelimb SEP reference signal (no injury) while $y_n$ is the hindlimb SEP signal corresponding to various degrees of injury. Note that the cardinal signal has been constructed to emulate the shape of the SEP signal over the regions of physiological significance,  while it is made more permissive toward the beginning and the end in order to deemphasize  transient/settling effects.  $K=7$ subranges are shown, as allowed by the proposed generalized ASCI. 

Figure \ref{ascik} shows the generalized ASCI values computed using (\ref{asci1}) for various choices of the number of subranges $K$. It can be seen that using a standard trichotomization with $K=3$ (corresponding to conventional ASCI)  does not yield ASCI values that are reflective of the signal similarity. { In particular, for small values of $K$, the Moderate, Severe I, and Severe II categories have very similar generalized ASCI values which fail to clearly differentiate between the injury categories. This is because the coarseness of the subranges causes the signals to be inappropriately classified, which can be accentuated by the effects caused by the transients at the beginning and end of the signal.} As the value of $K$ is increased, the distinction between these categories is enhanced. This is due to the improved resolution afforded by the larger value of $K$ (and hence the larger number of subranges) that is accommodated by the proposed generalized ASCI. It should also be noted that the value of the generalized ASCI stabilizes around $K=15$, indicating that this is the optimal number of subranges to select for the present scenario.

{The values of the spectral coherence method \cite{AlNashashBET0809} are shown in Figure \ref{ascik} for comparison. It can be seen that the spectral coherence values compare very favorably with the generalized ASCI for $K\geq 15$. However, it should be noted that computing the spectral coherence requires estimation of the power spectra of the forelimb and hindlimb SEP, and is significantly more computationally demanding than  the generalized ASCI. Indeed, the run time in MATLAB for computing the spectral coherence is around 50ms whereas it is only around 10ms for computing the generalized ASCI. This computational savings stems from the fact that the generalized ASCI does not involve complex terms and can be implemented with a minimal number of floating point operations, thus making it an attractive option for low-power, embedded medical applications.}

\section{Conclusion}
\label{sec6}

{A simple and computationally efficient method is proposed for assessing morphological similarity of two signals by generalizing the conventional ASCI to  an arbitrary number of  levels for resolution enhancement. A computational example was presented to demonstrate the ability of the proposed method to offer a more refined estimate of the similarity of SEP signals when assessing thoracic-level SCI. The results demonstrate that the proposed method yields similar results to the pioneering spectral coherence method, but at lower computational cost. Future work may include integrating the proposed method as part of a low-cost, low-power embedded system for the regular assessment of SCI.}

\vskip5pt

\noindent A. Olenko (\textit{La Trobe University, Australia})

\noindent K. T. Wong (\textit{Hong Kong Polytechnic University, Hong Kong})

\noindent H. Mir (\textit{American University of Sharjah, UAE})

\noindent H. Al-Nashash (\textit{American University of Sharjah, UAE})
\vskip2pt 

\noindent E-mail: ktwong@ieee.org

\end{document}